\documentclass{article}
\usepackage{amsmath,amsthm}
\usepackage{amssymb,latexsym}
\usepackage[mathscr]{eucal}
\usepackage{setspace}
\usepackage{graphics}
\usepackage{graphicx}
\graphicspath{{converted_graphics/}}
\usepackage{epsfig}
\usepackage{array}
\usepackage{epsdice}
\usepackage{epsf}
\usepackage{epstopdf}

\newcommand{\al}{\alpha}
\newcommand{\si}{\sigma}
\newcommand{\lam}{\lambda}

\newcommand{\pr}{\prime}

\newcommand{\lp}{\left(}
\newcommand{\rp}{\right)}
\newcommand{\lb}{\left[}
\newcommand{\rb}{\right]}
\newcommand{\lc}{\left\{}

\newcommand{\la}{\langle}
\newcommand{\ra}{\rangle}
\newcommand{\be}{\begin{equation}}
\newcommand{\ee}{\end{equation}}

\begin{document}

\begin{center}
\huge{\bf A Generalized Brownian Motion Model for Turbulent Relative Particle Dispersion}

\vspace{.3in}

\large{B.K. Shivamoggi\footnote{\normalsize Permanent Address: University of Central Florida, Orlando, FL 32816-1364}\\ J.M. Burgers Centre and Fluid Dynamics Laboratory\\
Department of Physics\\
Eindhoven University of Technology\\
5600 MB Eindhoven, The Netherlands\\ and \\International Centre for Theoretical Sciences (ICTS-TIFR),\\ TIFR Centre Building, IISc Campus\\ Bengaluru 560012, India}
\end{center}

\vspace{.3in}



\noindent\Large{\bf Abstract}\\

\large In this paper, a generalized Brownian motion model has been applied to describe the relative particle dispersion problem in more realistic turbulent flows. The fluctuating pressure forces acting on a fluid particle are taken to be a colored noise and follow a stationary process and are described by the Uhlenbeck-Ornstein model while it appears plausible to take their  correlation time to have a power-law dependence on the flow Reynolds number $R_e$, thus introducing a bridge between the Lagrangian quantities and the Eulerian parameters for this problem. This {\it ansatz} is in qualitative agreement with the possibility of a connection speculated earlier by Corrsin \cite{corr} between the white-noise representation for the fluctuating pressure forces and the large-$R_e$ assumption in the Kolmogorov \cite{Kol} theory for the 3D fully developed turbulence (FDT) as well as the argument of Monin and Yaglom \cite{Moy} and the result of Sawford \cite{saw1} and Borgas and Sawford \cite{bosa} that the Lagrangian acceleration is delta-function auto-correlated in the infinite-$R_e$ limit. It also provides an insight into the result that the Richardson-Obukhov scaling holds only in the infinite-$R_e$ limit and disappears otherwise. This {\it ansatz} further confirms the Lin-Reid \cite{linre} conjecture regarding the connection between the fluctuating pressure-force parameter and the energy dissipation rate in turbulence and leads to an $R_e$-dependent explicit relation between the two speculated by Lin and Reid \cite{linre}. More specifically, this {\it ansatz} provides a determination of the Richardson-Obukhov constant $g$ as a function of $R_e$, with an asymptotic constant value in the infinite-$R_e$ limit. It is shown to lead to full agreement, in the small-$R_e$ limit as well, with the Batchelor-Townsend \cite{Bat1} scaling for the rate of change of the mean square interparticle separation in 3D FDT, hence validating its soundness further.

\pagebreak

\noindent\Large{\bf 1. Introduction}\\

\large Taylor \cite{Tay} introduced the concept of turbulent diffusion by the continuous movement of a single particle and defined a diffusion coefficient $\mathscr D$ such that the mean square displacement $\la[ s(t) ]^2\ra$ is given by 

\be
\la[s(t)]^2\ra \ = 2\mathscr Dt
\ee

\noindent $t$ being the elapsed time\footnote{$t$ is assumed to be sufficiently large so the memory of the conditions of the initial release has been lost.}. The single-particle diffusion characterizes the translational motion of a small cloud released at $t=0$ in a turbulent flow while the rate of spreading of this cloud is measured by the average rate at which two particles moving with this cloud separate due to turbulent advection. On the other hand, the two particles may be expected to meander together under the influence of large-scale tranlating motions while they will drift apart under the influence of small-scale straining motions.  Richardson \cite{Rich}, therefore, proposed that turbulent diffusion should instead be characterized by the distance between neighboring particles, and hence by the effective shear that acts over the interparticle separation rather than the magnitude of the turbulent velocity field. If the interparticle distance is much greater than the integral scale $L$, the particles are uncorrelated and hence separate diffusively; the particles separate chaotically if the interparticle distance is much less than the Kolmogorov microscale $\eta$. On the other hand, if the interparticle distance is within the inertial range (where there is no characteristic length scale), the properties of turbulence are characterized by only one parameter - the mean energy dissipation rate $\varepsilon$. Consequently, one may then expect to find universal superdiffusive behavior in the relative particle dispersion process, which may be interpreted in terms of an interparticle-separation dependent turbulent relative diffusivity. From a purely empirical analysis of atmospheric dispersion data, Richardson \cite{Rich} then surmised that the turbulent relative diffusivity defined by the rate of increase of the mean square interparticle separation distance $R(t)$,
 
\be
\mathscr D \equiv \frac{1}{2} \frac{d}{dt} \la \lb R \lp t \rp \rb^2 \ra
\ee

\noindent goes like $4/3$ power of this distance, 

\be
\mathscr D \sim \la \lb R \lp t \rp \rb^2 \ra^{2/3}
\ee

\noindent implying that the turbulent relative dispersion is an accelerating process because, as the particles separate further, the relative dispersive-motion scales (bounded above by $R\lp t\rp$) become larger.

Obukhov \cite{Obu} showed that Richardson's relation (3) can be derived via Kolmogorov's \cite{Kol} theory for homogeneous isotropic 3D fully developed turbulence (FDT). When the interparticle separation is within the inertial range, Obukhov \cite{Obu} gave

\be
\mathscr D \sim \varepsilon^{1/3} \la \lb R \lp t \rp \rb^2 \ra^{2/3}
\ee

\noindent hence the name Richardson-Obukhov (RO) scaling. On the other hand, Richardson's \cite{Rich} formulation very nicely connects with the universal aspects of FDT since the unphysical effects due to sweeping by large scales $L$ are precluded from the outset in Richardson's \cite{Rich} formulation.

The RO scaling result has, however, remained elusive and received little experimental support due to the difficulty of performing Lagrangian measurements over a broad enough range of time and with sufficient accuracy. Even in recent laboratory experiments (Ott and Mann   \cite{OM}, Sawford  \cite{Saw}, Bourgoin et al. \cite{Bou}, Salazar and Collins  \cite{SaCo}, Sawford and Pinton \cite{sapi}) with high-speed photography to track particles and in numerical simulations (Yeung \cite{Ye}, Boffetta and Sokolov \cite{BoSo}) with the highest possible resolution for homogeneous isotropic turbulence, it is known to be hard to obtain an extended range with the RO scaling. Indeed, Bourgoin et al. \cite{Bou} reported that the RO scaling may not be observable in laboratory experiments even at high Reynolds numbers unless the initial particle separation is significantly small (so as to preclude the ballistic regime (Batchelor \cite{Bat2}) that currently seems to dominate the observable regime). The difficulty of achieving the RO scaling in laboratory experiments and numerical simulations appears to be due to,

\begin{itemize}
  \item[*] shrinking of the inertial range and enhancement of finite Reynolds number effects in the Lagrangian statistics (Sawford \cite{saw1}, Mordant et al. \cite{mor});
 \item[*] contamination of the inertial range by the usual dissipative effects at the ultraviolet end and by the external forcing effects at the infrared end of the spectrum caused by inadequate scale separation;
 \item[*] persistent memory of initial separation;\end{itemize}
 
\noindent and is therefore basically traceable to the finiteness of the Reynolds number as well as the limited observational domain. The purpose of this paper is therefore to try to shed some light on the former aspect by taking recourse  to the Lagrangian characterization of turbulent flows which shift the emphasis from spatial correlation of instantaneous velocity fields to temporal correlations along the path lines of fluid particles. More specifically, we consider a generalized Brownian motion model describe to the relative particle dispersion problem extending Obukhov's \cite{Obu2} idea further, for application to more realistic turbulent flows. The fluctuating pressure forces acting on a fluid particle are not taken to be a white-noise\footnote{In a real turbulent flow, the motion of a fluid particle, at any instant, cannot be taken to be uncorrelated with its motion at previous instants unlike a particle undergoing a Brownian motion (Batchelor \cite{Ba}).}  (considered by Lin \cite{lin}, Lin and Reid \cite{linre}) and the lack of "non-whiteness" is tied to the finiteness of the Reynolds number, thus introducing a bridge between the Lagrangian quantities and the Eulerian parameters for this problem.\footnote{General difficulties underlying the task of relating Lagrangian and Eulerian statistics quantities were emphasized by Lumley \cite{lum}.} \\ \\

\noindent\Large{\bf 2. Application of a Generalized Brownian Motion Model}\\

\large A fluid particle in a turbulent flow moves in response to fluctuations in the pressure of the surrounding fluid. The assumption of short-range correlation of the fluctuating pressure forces in a turbulent system motivates the applicability of a Brownian motion model to describe the turbulent dispersion problem (Obukhov \cite{Obu2}). The fluid particle is assumed to be subjected to successive small impulsive pressure forces, like those experienced by a Brownian particle. Obukhov \cite{Obu2} therefore proposed that the motion of a fluid particle mimics a Markov process and is described by a Fokker-Planck equation. These assumptions hold even more strongly for the relative motion of two particles in view of the negligible effect of large-scale motions in this process (Lin \cite{lin}, Lin and Reid \cite{linre}). 

Consider the relative motion of two particles released at an infinitesimal distance apart with almost the same velocity in an unbounded fluid in a state of stationary homogeneous turbulence. So, the two particles will not follow the same trajectory and will drift apart due to different initial accelerations. We then consider the statistical average over an ensemble of such pairs. This problem is governed by the following initial-value problem (Lin \cite{lin}, Lin and Reid \cite{linre}),

\be
\frac{dR}{dt} = V
\ee

\be
\frac{dV}{dt} + \beta V = \alpha \lp t \rp 
\ee

\noindent with,

\be
t = 0 ~: ~R \approx 0, ~V \approx 0.
\ee

\noindent Here, $\beta$ is a constant coefficient of viscous drag on the particles and $\alpha \lp t \rp$  is the fluctuating differential pressure force\footnote{Theory (Yaglom \cite{ya}) and numerical simulations (Vedula and Yeung \cite{ved}) showed that the strongest contribution to the particle acceleration comes from the local pressure gradient.} which is assumed to follow a stationary process.

The formal solution (because $\alpha \lp t\rp$ is not known with precision) of (6) and (7) for a particle pair is 

\be
V \lp t \rp = \int^t_0 e^{-\beta \lp t - \xi \rp} \alpha \lp \xi \rp d \xi.
\ee

The mean square of relative velocity of an ensemble of such particle pairs (satisfying the initial conditions (7)) is

\be
\la \lb V \lp t \rp \rb^2 \ra = e^{-2 \beta t} \int^t_0 d \eta \int^t_0 d \xi e^{\beta \lp \xi + \eta \rp} \la \alpha \lp \xi \rp \alpha \lp \eta \rp \ra.
\ee

We now generalize the Lin-Reid \cite{linre} Brownian motion model and assume that the fluctuating pressure force $\alpha \lp t \rp$ is a colored noise and has an auto-correlation function given by the Uhlenbeck-Ornstein \cite{uhor} model,

\be
\la \alpha \lp t^\pr \rp \alpha \lp t^{\pr \pr} \rp \ra = \sigma^2 e^{-\lam \lp t^\prime - t^{\pr \pr} \rp}
\ee

\noindent $\lam$ being the inverse of the correlation time of the fluctuating pressure forces. Large $\lam$ (the case considered by Lin \cite{lin} and Lin and Reid \cite{linre}) corresponds to white noise (the fluctuating pressure force is delta-function auto-correlated)\footnote{Since the Lagrangian acceleration is primarily determined by the Kolmogorov-microscale size motions (see Section 3), the Lagrangian acceleration auto-correlation function, in the inertial range at large Reynolds numbers, may be expected to be a delta function (Monin and Yaglom \cite{Moy}, p. 370). This was also confirmed by Sawford \cite{saw1} and Borgas and  Sawford \cite{bosa}. However, the white noise assumption for the fluctuating pressure forces in a real turbulent flow situation, at finite Reynolds numbers,  does not really seem to be justifiable, because real turbulent flows are spatially non-smooth but temporally finite-correlated (Falkovich and Frishman \cite{FalFrish}).} while small $\lam$ corresponds to persistent pressure fluctuation correlation. Interesting alternative interpretations for the parameter $\lam$ become possible upon further development with this model, as seen in the following.

Using (10), (9) becomes

\be
\la \lb V \lp t \rp \rb^2 \ra = \frac{\sigma^2}{\lam} \lb \frac{\lam}{\beta} \lp \frac{1}{\lam + \beta} - \frac{e^{-2 \beta t}}{\lam - \beta} \rp + \frac{2 \lam}{\lam^2 - \beta^2} e^{-\lp \beta + \lam \rp t} \rb.
\ee

For $\lp\lam/\beta\rp\gg1,$ (11) becomes (Lin and Reid \cite{linre}), 

\be
\la \lb V \lp t \rp \rb^2 \ra \approx \frac{\sigma^2/\lam}{\beta} \lp 1 - e^{-2 \beta t} \rp
\ee

\noindent which leads to

\begin{figure}
\begin{center}
\includegraphics[scale=.9]{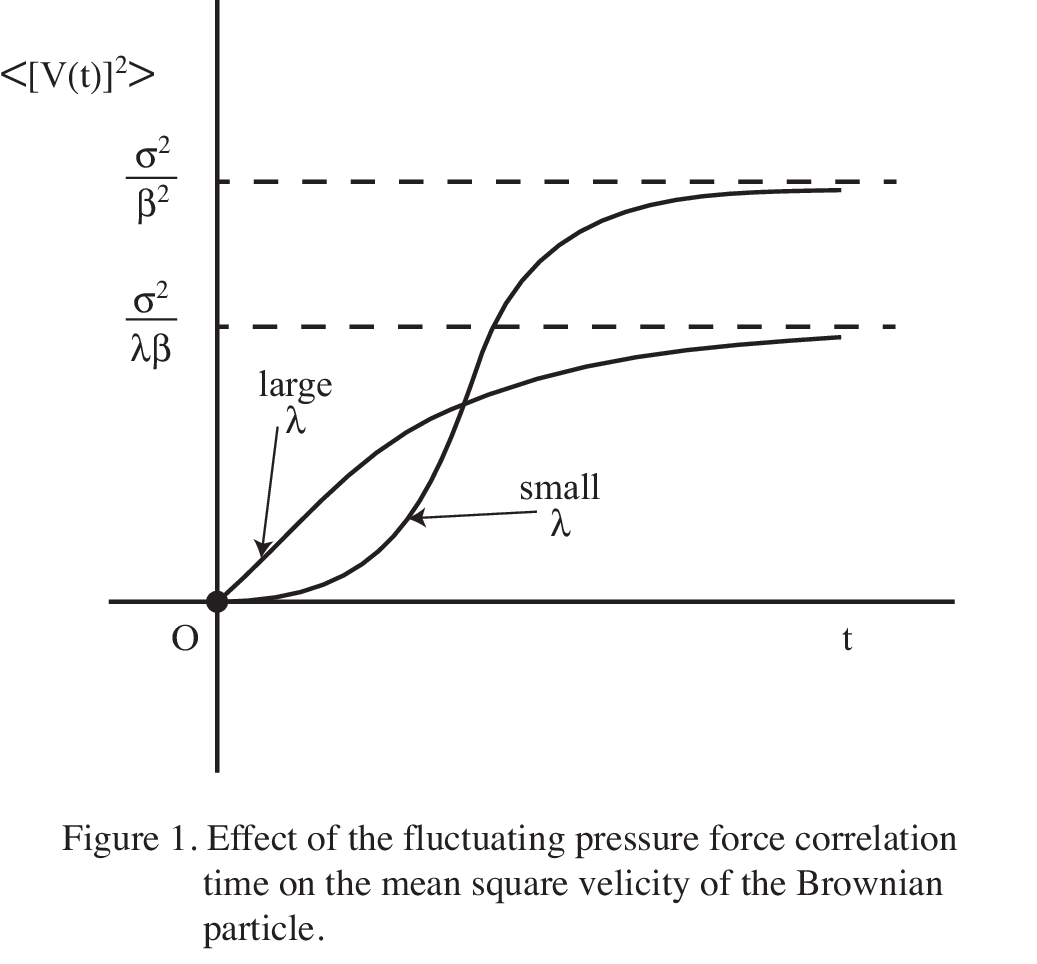}
\end{center}
\end{figure}

\be\tag{13a,b} 
\la \lb V (t) \rb^2 \ra \approx \lc
\begin{matrix}
\displaystyle 2 \lp \frac{\sigma^2}{\lam} \rp t, & \beta t \ll 1\\
\\
\displaystyle \frac{\lp \sigma^2/\lam \rp}{\beta}, & \beta t \Rightarrow \infty.
\end{matrix}
\right.
\ee

(12) or (13) shows that, in the absence of dissipation $\lp \beta = 0 \rp$, the particle pair experiences a runaway drifting motion due to the cumulative effect of the fluctuating pressure forces acting on the particle pair, which have been assumed to follow a stationary process.

For $\lp\lam/\beta\rp\ll 1$, on the other hand, (11) becomes

\be\tag{14}
\la \lb V \lp t \rp \rb^2 \ra \approx \lp \frac{\sigma^2}{\lam} \rp \frac{\lam}{\beta^2} \lp 1 + e^{-2 \beta t} - 2 e^{-\beta t} \rp
\ee

\noindent which leads to, 

\be\tag{15a,b}
\la \lb V \lp t \rp \rb^2 \ra \approx \lc
\begin{matrix}
\sigma^2 t^2, & \beta t \ll 1\\ 
\\
\displaystyle \frac{\sigma^2}{\beta^2}, & \beta t \Rightarrow \infty.
\end{matrix}
\right.
\ee

\noindent Comparison of (14) with (12) (see Figure 1) shows that the mean square of relative velocity initially grows more rapidly for the short-range correlation case than for the long-range correlation case.

Next, using (8), (5) and (7) give

\be\tag{16}
R \lp t \rp = \frac{1}{\beta} \int^t_0 \alpha \lp \xi \rp d \xi - \frac{1}{\beta} \int^t_0 e^{-\beta \lp t - \xi \rp} \al \lp \xi \rp d \xi.
\ee

The relative particle dispersion is then given by 

\be\tag{17}
\begin{aligned}
\la \lb R \lp t \rp \rb^2 \ra ~& = ~\frac{1}{\beta^2} e^{-2 \beta t}\int^t_0 d \eta \int^t_0 d \xi ~e^{\beta \lp \xi + \eta \rp} \la \al \lp \xi \rp \al \lp \eta \rp \ra\\ 
\\
& + \frac{1}{\beta^2} \int^t_0 d \eta \int^t_0 d \xi \la \al \lp \xi \rp \al \lp \eta \rp \ra - \frac{2}{\beta^2} e^{-\beta t} \int^t_0 d \eta \int^t_0 d \xi ~e^{\beta\xi} \la \al \lp \xi \rp \al \lp \eta \rp \ra.
\end{aligned}
\ee

Using (10), (17) becomes

\be\tag{18}
\begin{aligned}
\la \lb R (t) \rb^2 \ra & = \frac{\lp \sigma^2/\lam \rp}{\beta^2} \lb \frac{\lam}{\beta} \lp \frac{1}{\lam + \beta} - \frac{e^{-2 \beta t}}{\lam - \beta} \rp + \frac{2 \lam}{\lam^2 - \beta^2} e^{-\lp \lam + \beta \rp t} \rb\\
\\
& + 2 \frac{\lp \sigma^2/\lam \rp}{\beta^2} \lp t - \frac{1}{\lam} + \frac{1}{\lam} e^{-\lam t} \rp - 2 \frac{\lp \sigma^2/\lam \rp}{\beta^2} \lb \frac{2 \lam + \beta}{\beta \lp \lam + \beta \rp} + \frac{1}{\lp \lam + \beta \rp} e^{-\lp \lam + \beta \rp t} \right.\\
\\
& \left. + ~\frac{1}{\lp \lam - \beta \rp} e^{-\lam t} - \frac{2 \lam - \beta}{\beta \lp \lam - \beta \rp} e^{-\beta t} \rb.
\end{aligned}
\ee

For $\lp\lam/\beta\rp\gg 1$, (18) becomes (Lin and Reid  \cite{linre}), 

\be\tag{19}
\la \lb R \lp t \rp \rb^2 \ra \approx \frac{\sigma^2/\lam}{\beta^3} \lp 1 - e^{-2 \beta t} \rp + 2 \frac{\sigma^2/\lam}{\beta^2} t - 4 \frac{\sigma^2/\lam}{\beta^3} \lp 1 - e^{-\beta t} \rp\\
\ee

\noindent which leads to

\be\tag{20a,b}
\la \lb R \lp t \rp \rb^2 \ra \approx \lc
\begin{matrix}
\displaystyle \frac{2}{3} \lp \frac{\sigma^2}{\lam} \rp t^3, ~\beta t \ll 1\\
\\ 
2 \displaystyle \frac{\lp \sigma^2/\lam \rp}{\beta^2} t, ~\beta t \Rightarrow \infty.
\end{matrix}
\right.
\ee
 
(20a) represents the Richardson \cite{Rich} scaling and signifies the validity of a Brownian motion theoretical framework underlying the relative particle dispersion problem (Lin \cite{lin}, Lin and Reid \cite{linre}). Indeed, using (2) and (20a), the turbulent relative diffusivity is given by (Lin and Reid \cite{linre})

\be\tag{21}
D = \lp \frac{\sigma^2}{\lam} \rp t^2 = \lp \frac{3}{2} \rp^{2/3} \lp \frac{\sigma^2}{\lam} \rp^{1/3} \lp \la \lb R \lp t \rp \rb^2 \ra \rp^{2/3}
\ee

\noindent which is Richardson's \cite{Rich} other scaling result.

In view of the consistency of Richardson \cite{Rich} scaling with Kolmogorov \cite{Kol} scaling (which is valid in the infinite flow Reynolds number limit), it therefore appears plausible to surmise that the Brownian parameter $\lp\lam/\beta\rp$ is an increasing function of the flow Reynolds number. One may assume, for the sake of illustration, the simplest \textit{ansatz}, for which $\lp\lam/\beta\rp$ has a power-law dependence on the flow Reynolds number $R_e$, 

\be\tag{22}
\lam/\beta=k R^\delta_e,~~ \delta>0
\ee

\noindent where $k$ is a constant and,

\be\tag{23}
R_e\equiv\frac{Ua}{\nu}
\ee

\noindent $U$ being a reference particle speed, $a$ being the particle radius and $\nu$ being the viscosity of the fluid. On estimating on dimensional grounds, in the large-$R_e$ limit, that

\be\tag{24}
\beta\sim\frac{\nu}{a^2}
\ee
\noindent (22) yields,

\be\tag{25}
\frac{\lam~ a}{U}\sim R_e^{\lp\delta -1\rp}.
\ee

\noindent (25) implies a more refined lower bound on $\delta$\footnote{Experimental data (Mordant et al. \cite{mor}) indicated that the particle-acceleration statistics has a structure with a width proportional to $R_e^{-1/2}$ (which would correspond to $\delta\approx3/2$).}:

\be\tag{26} 
\delta >1
\ee

\noindent so the white-noise model for the fluctuating pressure forces corresponds to the infinite-$R_e$ limit. 

\indent{\it Ansatz} (22) introduces a bridge between the Lagrangian quantities and the Eulerian parameters for this problem. The importance of such a connection on a qualitative level was emphasized earlier by Lin \cite{lin} (see Section 3).  It is of further interest to note that the possibility of a connection between the white-noise representation for the fluctuating pressure forces and the large-$R_e$ assumption in Kolmogorov's \cite{Kol} theory for 3D FDT indicated by (22) was indeed speculated earlier by Corrsin \cite{corr}. On the other hand, Monin and Yaglom \cite{Moy} (p. 548)  expressed skepticism that the Lin-Reid \cite{linre} result (13) seemed not to demand that $R_e$ is large enough to validate it. Sawford \cite{saw1} and Borgas and Sawford \cite{bosa} also showed that the Lagrangian acceleration is delta-function auto-correlated in the infinite-$R_e$ limit. It is of further interest to note that (22), on the other hand, turns out to lead to full agreement, in the small-$R_e$ limit, with the Batchelor-Townsend \cite{Bat1} scaling for the rate of change of the mean square interparticle separation in 3D FDT (see Section 3), hence further validating its soundness.

For $\lp\lam/\beta\rp \ll 1$, (18) becomes

\be\tag{27}
\la \lb R \lp t \rp \rb^2 \ra \approx 2 \frac{\lp \si^2/\lam \rp}{\beta^2} t + \frac{\si^2}{\lam} \frac{\lam}{\beta^4} \lp 1 + e^{-2 \beta t} - 2 e^{-\beta t} \rp
\ee

\noindent which leads to

\be\tag{28a,b}
\la \lb R \lp t \rp \rb^2 \ra \approx \lc
\begin{matrix}
\displaystyle \frac{\lp \si^2/\lam \rp}{4} \lam t^4, ~\beta t \ll 1\\ 
\\
2 \displaystyle \frac{\lp \si^2/\lam \rp}{\beta^2} t, ~\beta t \Rightarrow \infty.
\end{matrix}
\right.
\ee

\noindent(22) and (28a) show that the Richardson \cite{Rich} scaling holds only in the infinite-$R_e$ limit and disappears at finite $R_e$'s. The turbulent relative diffusivity $D$, defined as per (2), is now given by

\be\tag{29}
D = \frac{1}{2} \lp \frac{\si^2}{\lam} \rp \lam t^3 = \lb 4 \lam \lp \frac{\si^2}{\lam} \rp \rb^{1/4} \lp \la \lb R \lp t \rp \rb^2 \ra \rp^{3/4}
\ee

\noindent which shows that the turbulent relativity diffusivity $D$, in the small-$R_e$ limit, increases as the $3/2$ power of the interparticle separation distance. (20b) and (28b), on the other hand, confirm that the relative dispersion process becomes diffusive, as to be expected, in the large-$t$ limit, irrespective of whether the fluctuating pressure forces are short-range or long-range correlated.

\vspace{.3in}

\noindent\Large{\bf 3. Relation between the Lagrangian Quantities and the Eulerian Parameters}\\

\large In the generalized Brownian motion framework, as indicated by (18), the relative particle dispersion depends on the fluctuating pressure-force parameter $\lp \si^2/\lam \rp$. Comparison of (21) with (4) shows that the Lagrangian parameter $\lp\si^2/\lam\rp$ looks very much like the energy dissipation rate parameter $\varepsilon$ in the Eulerian framework ($\si^2/\lam$ and $\varepsilon$ even have the same dimensions). Lin and Reid \cite{linre} therefore went on to contemplate the possibility that ``$\lp \si^2/\lam ~\text{and} ~\varepsilon \rp$ {\it might be identifiable, apart from a constant factor. But, in general, one can only surmise that perhaps} ({\it the Richardson-Obukhov constant}, $g\equiv$) $\displaystyle \frac{\si^2/\lam}{\varepsilon} = f \lp R_e \rp$." Lin \cite{lin} further emphasized that determination of the function $f(R_e)$ would be {\it "a critical step in connecting the Eulerian and the Lagrangian descriptions"}.  We will now show that \textit{ansatz} (22) indeed underscores such a connection and hence lends credence to the Brownian motion underpinnings for the Kolmogorov \cite{Kol} theory.

Multiplying equation (6) by $V$ and taking the average over an ensemble of such particle pairs, we obtain

\be\tag{30}
\frac{1}{2} \frac{d}{dt} \la \lb V \lp t \rp \rb^2 \ra + \beta \la \lb V \lp t \rp \rb^2 \ra = \la \al \lp t \rp V \lp t \rp \ra.
\ee

\noindent Upon substituting (8) and using (10) on the right in (30), we obtain

\be\tag{31}
\frac{1}{2} \frac{d}{dt} \la \lb V \lp t \rp \rb^2 \ra + \beta \la \lb V \lp t \rp \rb^2 \ra = \frac{\lp \si^2/\lam \rp}{1 + \lp \beta/\lam \rp} \lb 1 - e^{-\lp \beta + \lam \rp t} \rb.
\ee

\noindent Further, on assuming a stationary state in the limit $t \rightarrow \infty$, (27) yields 

\be\tag{32}
\varepsilon \sim \beta \la \lb V \lp t \rp\rb^2 \ra = \frac{\lp \si^2/\lam \rp}{1 + \lp \beta/\lam \rp},
\ee

\noindent which may be viewed as a statement of the {\it fluctuation-dissipation theorem}. 

On using \textit{ansatz} (22), (32) leads to 

\be\tag{33}
g \equiv\frac{\lp \si^2/\lam \rp}{\varepsilon} = f \lp R_e \rp \sim 1 + \lp \frac{1}{k R^\delta_e} \rp
\ee

\noindent which confirms the Lin-Reid \cite{linre} conjecture and further provides one with an explicit determination of $f \lp R_e \rp$ hinted at by Lin and Reid \cite{linre}. (33) also shows that $\lp\sigma^2/\lam\rp$ and $\varepsilon$ are proportional to each other in the infinite-$R_e$ (or white-noise) limit, and in general, the Richardson-Obukhov constant $g$ is a function of $R_e$, with an asympototic constant value in the infinite-$R_e$ limit - the latter aspect seems to be in agreement with the statistical-scaling based result of Franzese and Cassiani \cite{frca}\footnote{The numerical value of the Richardson-Obukhov constant $g$ remains controversial, due to considerable uncertainity in its estimate through atmospheric measurements, laboratory experiments and numerical simulations. The qualitative trend exhibited by laboratory-experiment and numerical-simulation data on the $R_e$-dependence of $g$ is also inconclusive (Franzese and Cassiani \cite{frca}).}.

Using (33), the small-$R_e$ result (29), on the other hand, becomes

\be\tag{34}
D \sim \left[ 1 + \left( \frac{1}{k R^\delta_e}\right)\right] \varepsilon (k\beta R^\delta_e) t^3 .
\ee

\noindent or

\be\tag{35}
D \sim (\beta \varepsilon) t^3 .
\ee

\indent Now, in the spirit of (22), one may conjecture a relation, in the small-$R_e$ limit, between the other Lagrangian quantity $\beta$ and the Eulerian parameters $\varepsilon$ and $\nu$ (which is the viscosity of the fluid), and estimate on dimensional grounds,

\be\tag{36}
\beta \sim \frac{\nu}{\eta^2}
\ee

\noindent $\eta$ being the Kolmogorov microscale,

\be\tag{37}
\eta \sim \left( \frac{\nu^3}{\varepsilon} \right)^{1/4} .
\ee

Using (37), (36) becomes

\be\tag{38}
\beta \sim \frac{\varepsilon^{1/2}}{\nu^{1/2}} .
\ee

Using (38), (35) becomes

\be\tag{39}
D \sim \frac{\varepsilon^{3/2}}{\nu^{1/2}} t^3 
\ee

\noindent which reflects the Heisenberg \cite{hci}-Yaglom \cite{ya} scaling for the Lagrangian acceleration statistics. In fact, using (29) and (37), (39) yields for the particle-acceleration variance,

\be\tag{40}
\sigma^2\sim\varepsilon^{4/3}\eta^{-2/3}
\ee

\noindent which agrees with the dissipation-range scaling of the Lagrangian acceleration variance according to the Kolmogorov similarity theory (Sawford \cite{saw1}) and the scaling behavior of the second-order structure function of the pressure gradients (Batchelor \cite{bat3}), and the pressure (as per laboratory experiment (Ould-Rouis et al. \cite{Or}), and numerical simulation (Gotoh and Fukuyama \cite{Gf})), extrapolated to the Kolmogorov microscale\footnote{This extrapolation is rationalized by the insensitivity of the scaling behavior of the pressure structure function to variation of the Reynolds number (Batchelor \cite{bat3}) On the other hand, the Heisenberg-Yaglom scaling, reflected in (39), actually implies that the particle-acceleration variance is determined primarily by the Kolmogorov-microscale size motions (Monin and Yaglom \cite{Moy}, p.370).}.

\indent On the other hand, in the Eulerian framework\footnote{In the Eulerian framework, the short-time relative particle dispersion is found to be constrained by some compatibility conditions (Falkovich and Frishman \cite{FalFrish}).}, Batchelor and Townsend \cite{Bat1} used dimensional arguments to postulate that the rate of change of the mean square interparticle separation in 3D FDT is given by

\be\tag{41}
\frac{d}{dt} \la[R(t)]^2\ra \sim \varepsilon t^2 \cdot \tilde{f}\left( \frac{R(0)}{\varepsilon^{1/2} t^{3/2}} \ , \ \frac{\varepsilon^{1/2} t}{\nu^{1/2}} \right).
\ee

In the small-$R_e$ limit, on assuming,

\be\tag{42}
 \tilde{f}\left( \frac{R(0)}{\varepsilon^{1/2} t^{3/2}} \ , \ \frac{\varepsilon^{1/2} t}{\nu^{1/2}} \right) \sim \frac{\varepsilon^{1/2} t}{\nu^{1/2}}
\ee

\noindent (41) leads to

\be\tag{43}
D\sim\frac{d}{dt} \la[R(t)]^2\ra \sim \frac{\varepsilon^{3/2}}{\nu^{1/2}} t^3
\ee

\noindent in agreement with (39), hence providing further validation for the soundness of the empirical \textit{ansatz} (22).\\

\indent On the other hand, in the large-$R_e$ limit, the influence of fluid viscosity $\nu$ vanishes; (41) then leads to

\be\tag{44}
D\sim\frac{d}{dt} \la [R(t)]^2 \ra \sim \varepsilon t^2 \sim \varepsilon^{1/3} \la [R(t)]^2 \ra^{2/3}
\ee

\noindent in agreement with (4), as expected.
\vspace{.3in}

\noindent\Large{\bf 4.  Discussion}\\

\large There is speculation that the difficulty in obtaining an extended range with RO scaling in both laboratory experiments and numerical simulations is due to the finiteness of the flow Reynolds number $R_e$ in these situations. In this paper, a generalized Brownian motion model has been used to describe the relative particle dispersion problem in more realistic turbulent flows and to shed some light on this issue. The fluctuating pressure forces acting on a fluid particle are taken to be a colored noise and follow a stationary process and are described by the Uhlenbeck-Ornstein model while it appears plausible to take their correlation time to have a power-law dependence on $R_e$, thus introducing a bridge between the Lagrangian quantities and the Eulerian parameters for this problem (the importance of which on a qualitative level was emphasized earlier by Lin \cite{lin}). This $\textit{ansatz}$ is in qualitative agreement with the possibility of a connection speculated earlier by Corrsin \cite{corr} between the white-noise representation for the fluctuating pressure forces and the large-$R_e$ assumption in the Kolmogorov \cite{Kol} theory for 3D FDT as well as the argument of Monin and Yaglom \cite{Moy} and the  result of Sawford \cite{saw1} and Borgas and Sawford \cite{bosa} that the Lagrangian acceleration is delta-function auto-correlated in the infinite-$R_e$ limit. It also provides an insight into the result that the RO scaling holds only in the infinite-$R_e$ limit and disappears otherwise. This $\textit{ansatz}$ further confirms the Lin-Reid \cite{linre} conjecture regarding the connection between the fluctuating pressure-force parameter and the energy dissipation rate in FDT and leads to an $R_e$-dependent explicit relation between the two speculated by Lin and Reid \cite{linre} (the determination of which, according to Lin \cite{lin}, is {\it``a crucial step connecting the Eulerian and the Lagrangian descriptions"}). More specifically, this {\it ansatz} provides a determination of the Richardson-Obukhov constant $g$ as a function of $R_e$, with an asymptotic constant value in the infinite-$R_e$ limit. It turns out further to lead to full agreement, in the small-$R_e$ limit as well, with the Batchelor-Townsend \cite{Bat1} scaling for the rate of change of the mean square interparticle separation in 3D FDT, hence further validating its soundness further.

\vspace{.3in}

\noindent\Large{\bf Acknowledgments}\\

\large Much of this work was carried out during the course of a visiting research appointment at the Eindhoven University of Technology and I would like to thank The Netherlands Organization for Scientific Research (NWO) for the financial support. I am thankful to Professor Gert Jan van Heijst for his hospitality as well as discussions. Part of the work was carried out during my visiting research appointment at the International Centre for Theoretical Sciences, Bengaluru, India. I am thankful to Professor Spenta Wadia for his hospitality. I am thankful to Professor Katepalli Sreenivasan for his constant encouragement and helpful remarks. I am also thankful to Professors Grisha Falkovich and Predhiman Kaw for their valuable remarks.

\end{document}